\documentclass[aps,pre,twocolumn,superscriptaddress,floatfix,notitle]{revtex4-1}
\usepackage{graphicx,amsmath,amssymb,verbatim,color}
\newcommand{\be}{\begin{equation}}
\newcommand{\ee}{\end{equation}}

\begin{document}
\title{Multiple resource demands and viability in multiplex networks}
\author{Byungjoon Min}
\affiliation{Department of Physics, Korea University, Seoul 136-713, Korea} 
\author{K.-I. Goh}
\email{kgoh@korea.ac.kr}
\affiliation{Department of Physics, Korea University, Seoul 136-713, Korea} 
\date{\today}
\begin{abstract}
Many complex systems demand manifold resources to be supplied from distinct
channels to function properly, i.e, water, gas, and electricity for a city. 
Here, we study a model for viability of such systems demanding more than one type of vital resources 
produced and distributed by resource nodes in multiplex networks. 
We found a rich variety of behaviors such as discontinuity, bistability, and hysteresis in the fraction of viable nodes
with respect to the density of networks and the fraction of resource nodes.
Our result suggests that viability in multiplex networks 
is not only exposed to the risk of abrupt collapse but also suffers excessive complication in recovery.
\end{abstract}
\maketitle

Our life in modern society relies on interrelated infrastructure networks including
water supply networks, gas supply networks, and power grid systems \cite{rinaldi}.
For this reason, the network of connections between public utilities and consumers has
been studied for a long time~\cite{utility}.
Over the last decade, there have been a series of attempts 
to understand robustness of networks including infrastructure networks using their connectivity 
properties
in physics and network science communities~\cite{albert,callaway,cohen,holme,latora}.
Such studies on individual networks, however, could not fully assess the vulnerability of 
real infrastructure systems because many real systems are multiplex and interdependent \cite{rinaldi,rosato,buldyrev,dsouza}.
Moreover these systems often demand multiple classes of resources to be 
supplied through different layers of networks \cite{rinaldi}.
As a result, the vulnerability of interdependent infrastructure systems 
can be far beyond that expected in 
an individual network because the damage in one layer is not localized therein but
able to provoke avalanche collapse leading to an abrupt breakdown of the whole system \cite{buldyrev,li,bashan,son,baxter,weak}.

Most previous studies on interdependent and multiplex networks have assumed that mutual 
connectivity, that is the simultaneous connectivities through each and every network layer, is 
the requirement for an active node~\cite{buldyrev,son,baxter}.
Such condition is plausible for the networks in which connections by itself can provide function,
such as road networks and Internet.
This condition, however, is not sufficient for the systems 
in which the resource nodes generate products and distribute them along the links to the neighbors,
such as the power grid and the water supply networks.
For such kind of systems, simultaneous connectivities with resource nodes 
such as power plants in power grid and water sources in water supply networks 
through a series of functioning nodes are essential for the proper functioning, or to be {\em viable} as we will call it.
In this paper, we introduce and study a simple model of viability of multiplex systems requiring supports of more than one type of vital resources produced from a fixed set of resource nodes.
By presenting algorithms to identify the set of viable nodes and the analytic solutions to the problem,
we illustrate novel features of system behaviors characterized by the multiple resource demands.

We found that the final fraction of viable nodes exhibits discontinuous jumps, bistability, and hysteresis with the density of links and resource nodes.
The discontinuous jumps indicate a potential danger of abrupt collapse of 
the system similar to the previous study of cascading failures in interdependent networks~\cite{buldyrev}. 
It is noteworthy that in our model the discontinuity is still observed 
even though all the viable clusters or islands, rather than the giant viable component by itself, are considered.
Furthermore, the strong hysteresis with respect to the link density
suggests that after collapse, a far more addition of links compared with the link density before collapse is required to restore viability of networks to the level before collapse.
We also examine the effect of the number of the resource layers 
and find the expression for the critical point at which the 
discontinuity disappears for $n$-layer Erd\H{o}s-R\'enyi (ER) multiplex networks.
In addition, our model can also be interpreted as a unifying model of mutual percolation ~\cite{buldyrev,son,baxter} 
and cooperative epidemics on multiplex networks~\cite{janssen,bizhani,chen}.

Consider a network with $n$-multiple layers, where each layer of the network corresponds to 
a certain infrastructural network.
A given fraction, $\rho$, of resource nodes, located randomly, generates and distributes resources 
essential to be viable.
A key assumption of our model is that only viable nodes can function properly and transmit resources further to their connected neighbors.
Then, a node is viable only if it can reach, via the viable nodes, to a resource node in each and every layer.
We present two algorithms to identify the set of viable nodes, called the
cascade of activations (CA) and deactivations (CD).
An example of the CA and the CD algorithms applied to a multiplex network with two layers is illustrated in Fig.~1.
For the CA algorithm, initially all nodes except resource nodes are unviable.
Each step, unviable nodes that are linked to viable nodes through each and every layer
(colored green in Fig.~1) are activated to become viable. This cascade of activations continues until no nodes newly become viable.
For the CD algorithm, initially all nodes including resource nodes are viable.
Each step, the nodes that do not reach the resource nodes in any layers (colored green in Fig.~1) are deactivated to become unviable.
This cascade of deactivations continues until there are no nodes to deactivate.
In the limit $N \rightarrow \infty$, the final fraction of viable nodes $V$ is called the viability of the system.
Note that the viability $V$ obtained from the CA and the CD algorithms can be in general different from each other:
the viable nodes for the CA are always viable for the CD as well, but not {\it vice versa}.
Note also that in our model the capacity of source node is assumed to be unlimited, so that a single source node is able to generate sufficient amount of resources for the entire network.

\begin{figure}
\includegraphics[width=0.9\linewidth]{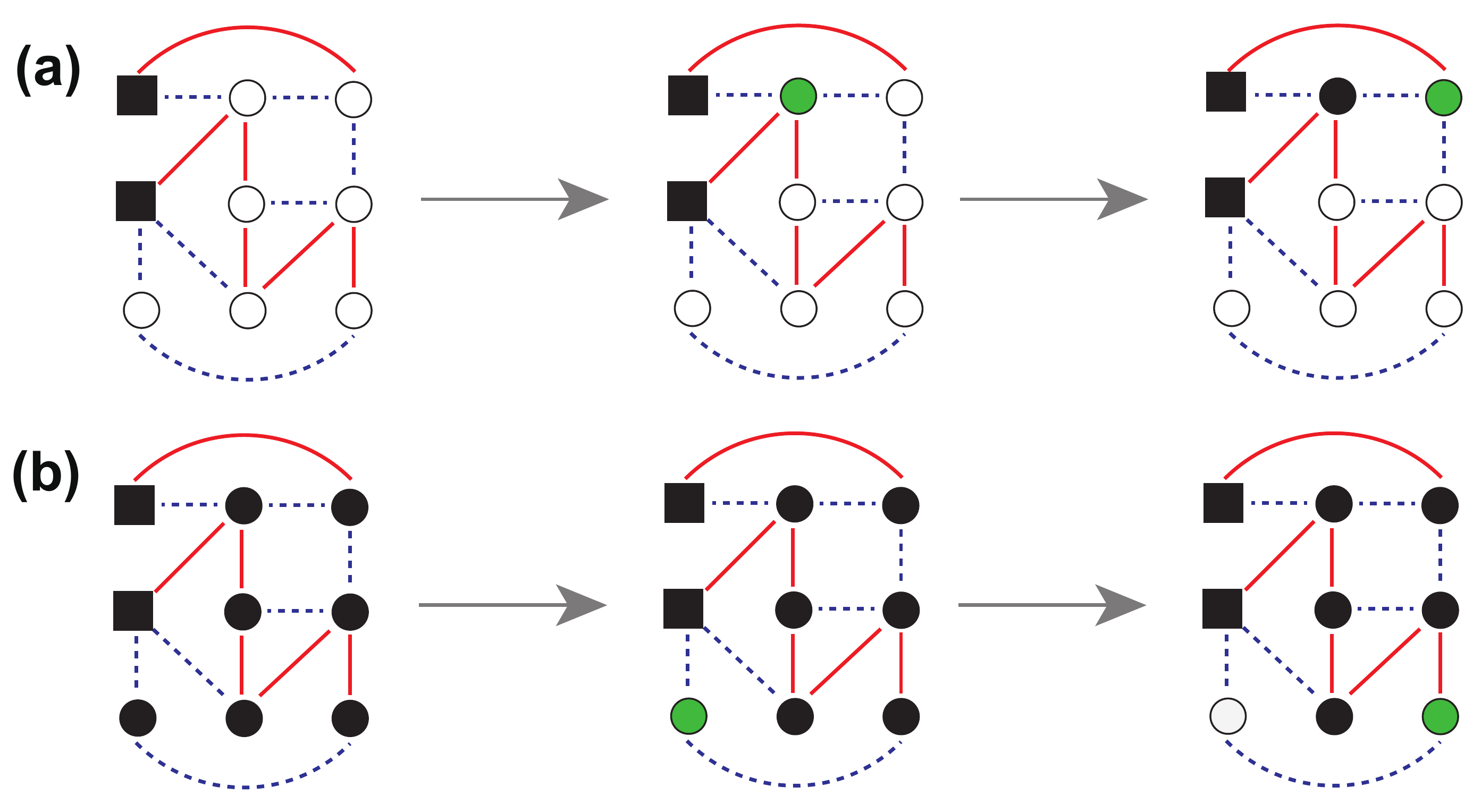}
\caption{(color online)
Illustrations of the iterative algorithms of (a) the cascade of activations (CA) and (b) deactivations (CD).
Source nodes (squares) generate resources. If a node connects with resource nodes 
through each type of links denoted by solid and dashed lines, the node is viable (filled circles) and
can transmit resources further to its neighbors. If not, the node is unviable (open circles).
Shaded (green) circles denote the node whose state is to be updated [activated in (a) and deactivated in (b)] at each step.
}
\end{figure}

One can also compute the viability analytically for a locally-tree-like structure, as follows. 
To calculate viability, we first consider the probability $u_i$ 
that a randomly chosen node reached by following an $i$-type link is not viable.
Each node has $\vec{k}=(k_1,\dots,k_n)$ degrees over different layers drawn from 
the joint degree distribution $p(\vec{k})$.
Given the initial fraction $\rho$ of randomly distributed resource nodes on locally-tree like structures
$u_i$ can be expressed as the self-consistency equations,
\begin{eqnarray}
1- u_i=\rho+(1-\rho)\sum_{\vec{k}} \frac{k_i p(\vec{k})}{z_i} (1-u_i^{k_i-1}) {\prod_{\substack{j=1 \\ j\ne i}}^n} (1-u_j^{k_j}),~~~~
\end{eqnarray}
where $z_i$ is the mean degree of the $i$-layer network.
The first term is the probability that a randomly chosen node is a resource node.
And the second term is the probability that a node is 
connected with viable nodes through each type of links.
The mean final fraction of viable nodes $V$, the viability, is the same as the
probability that a randomly chosen node is viable.
Therefore, $V$ can be similarly expressed as
\begin{eqnarray}
V=\rho+(1-\rho)\sum_{\vec{k}} p(\vec{k}) \prod_i (1-u_i^{k_i}).
\end{eqnarray}
By solving Eqs.~(1--2) with given $p(\vec{k})$ and $\rho$, one obtains the viability $V$. 

\begin{figure}[t]
\includegraphics[width=0.9\linewidth]{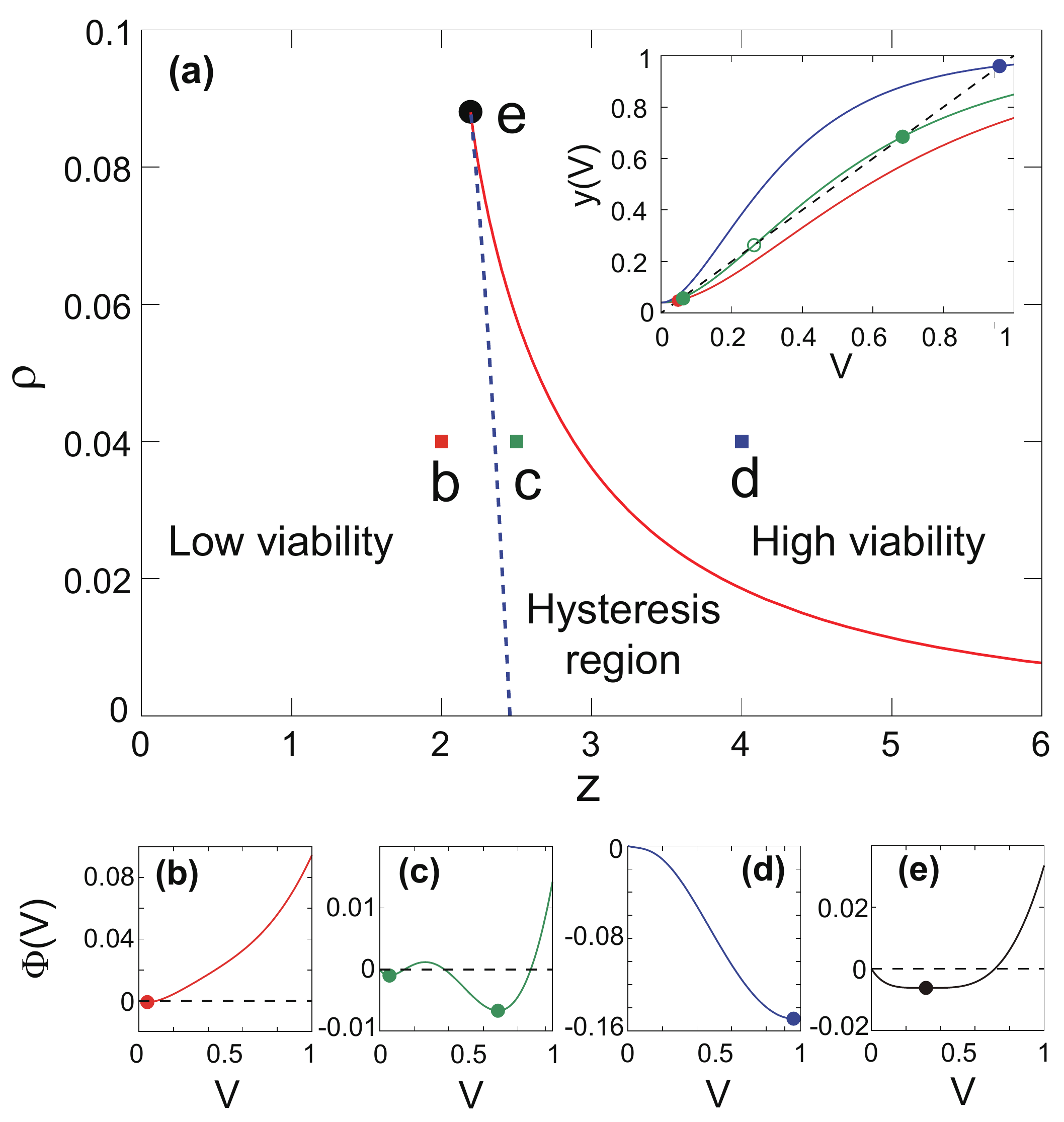}
\caption{
(a) Phase diagram of viability with respect to $\rho$ and $z$. 
Solid and dashed lines indicate the locations of a discontinuous jump 
for the CA and the CD algorithm, respectively.
Filled circle  
indicates the critical point at which the discontinuous jumps disappear.
(inset) Graphical solutions of Eq.~(3) for the viability of two-layer ER networks with equal mean degree $z$.
Solid lines indicate $y(V)=\rho+(1-\rho)(1-e^{-z V})^2$ with $z=2, 2.5, 4$ 
from bottom to top at a given $\rho=0.04$, and dashed line indicates $y(V)=V$.
Intersections of solid and dashed lines indicate
solutions either stable (filled circles) or unstable (open circles).
(b-e) Potential function at (b) $(\rho,z)=(0.04, 2)$, (c) $(0.04, 2.5)$,
(d) $(0.04, 4)$, and (e) $(\rho_c, z_c)$.
Local minima which give the viability $V$ are denoted by filled circles.
}
\end{figure}

We illustrate the basic features of the model with a specific example of a randomly-coupled multiplex network with two 
Erd\H{o}s-R\'enyi (ER) layers.
For simplicity we take the mean degrees of two layers to be the same, denoted as $z$,
and then Eqs.~(1) and (2) reduce to
\begin{eqnarray}
V=\rho+(1-\rho) (1-e^{-z V})^2.
\end{eqnarray}
By solving Eq.~(3) graphically, $V$ can be obtained
for given $\rho$ and $z$.
The typical behaviors of the graphical solution of Eq.~(3) with $\rho=0.04$
and $z=2$, $2.5$, and $4$ are shown in the inset of Fig.~2(a).
There is one stable solution for $z=2$ and $4$, corresponding to low ($V\approx0.05$) and high viability ($V\approx0.96$), respectively. In these cases, the outcomes of the CA and the CD algorithms coincide, being identical to $V$.
On the other hand, when $z=2.5$, there are two stable solutions, $V\approx0.06$ and $0.68$, corresponding to
the CA and the CD cases, respectively.
Such a bistability implies that the viability of networks is dependent on the 
initial fraction of viable nodes.
The bistability region is bounded by the saddle node bifurcations at $z_{\text{CD}}\approx 2.35$ and
 $z_{\text{CA}}\approx 2.88$ for $\rho=0.04$.

In the $(\rho,z)$ phase diagram [Fig.~2(a)], the low and high viability phases are separated by two lines, indicating the loci of saddle node bifurcations.
Defining $h(V)=V-\rho-(1-\rho)(1-e^{-z V})^2$, 
the locations of saddle node bifurcations can be determined by imposing the conditions, $h(R)=h'(R)=0$.
The solid line corresponds to the points at which the viability for the CA algorithm undergoes a discontinuous change, and the dashed line indicates that for the CD algorithm.
Between two lines (hysteresis region), there are two possible stable solutions, and $V$ is determined by its initial value.
For example, for the CA algorithm, $V$ keeps to be in low viable state until the abrupt jump at the solid line.
For the CD case, high viability sustains until abrupt collapse at the dashed line.
Two lines merge at the critical point located at 
$(\rho_c,z_c)=(\frac {2 \log 2-1}{2 \log 2 +3},\log 2 +\frac{3}{2})$
derived by the conditions, $h(V)=h'(V)=h''(V)=0$.
Above $\rho_c$, as $z$ increases, the viability changes gradually without discontinuity.
A similar phase diagram was found recently in 
a model of cooperative contagion with node recoveries 
showing spontaneous phase flipping behavior~\cite{majdandzic}.

For more intuitive understanding of the behavior of $V$, 
we define a potential function $\Phi(V)=-\int_0^V h(x) dx$.
For the randomly coupled two ER networks, the potential function
is obtained as 
\begin{equation}
\Phi(V)=-V+\frac{V^2}{2}-\frac{1-\rho}{2z}(4 e^{-zV}-e^{-2 zV}-3).
\end{equation}
$V$ descends along $\Phi(V)$ from its initial value,
and so finally remains at local minima
which are stable solutions of Eq.~(3) [Fig.~2(b-e)].
In particular, in the hysteresis region [Fig.~2(c)], 
two local minima in double wells correspond to the CA and CD algorithm, respectively,
and the local maximum corresponds to an unstable fixed point.

The bistability implies a hysteresis in viability of multiplex networks [Fig.~3(a)].
In order to demonstrate the hysteresis explicitly, 
let us suppose the following scenario of a sequence of systemic collapse 
and subsequent recovery of viability.
Initially the system is in high viability state with well-established networks.
As $z$ decreases by random failures of links, $V$ abruptly collapses at $z_{\text{CD}}$.
After collapse, if we try to restore viability to the level before the collapse,
more addition of links is needed up to $z_{\text{CA}}$ which is much larger than the point of collapse $z_{\text{CD}}$.
Thus, the bistability induces the hysteresis with $z$
which hinders and complicates recovery from the low viability state.
Our result shows that multiple resource demands produce not only a 
potential danger of abrupt collapse but also severe complication in recovery.
Also, note that the hysteresis does not occur in single networks, $n=1$.

To manage the potential risk of collapse and complication in recovery,
there can be two possible strategies:
More suppliers and denser networking.
First, as the number of resource nodes increases, meaning that $\rho$ increases, 
the gap between $z_{\text{CA}}$ and $z_{\text{CD}}$ decreases 
and eventually disappears for above $\rho_c$ as shown in [Fig.~3(b)].
Therefore, by placing enough resource nodes, i.e, $\rho>\rho_c$, one can avoid both discontinuity and hysteresis.
Another way to maintain high viability is dense networking.
When $z_c<z<z_{\text{MP}}$ where $z_c=\log 2+3/2\approx2.193$ and 
$z_{\text{MP}} \approx 2.455$, the threshold of mutual percolation for 2-layer ER networks \cite{buldyrev,son},
$V$ as a function of $\rho$ shows discontinue jumps and the hysteresis [Fig.~3(c)].
In case $z>z_{\text{MP}}$, however, the high viability corresponding to the CD algorithm is always guaranteed as long as $\rho\ne0$.
Thus, the network can maintain a highly viable state for $z>z_{\text{MP}}$,
even with extremely small density of resources.

\begin{figure}
\includegraphics[width=0.9\linewidth]{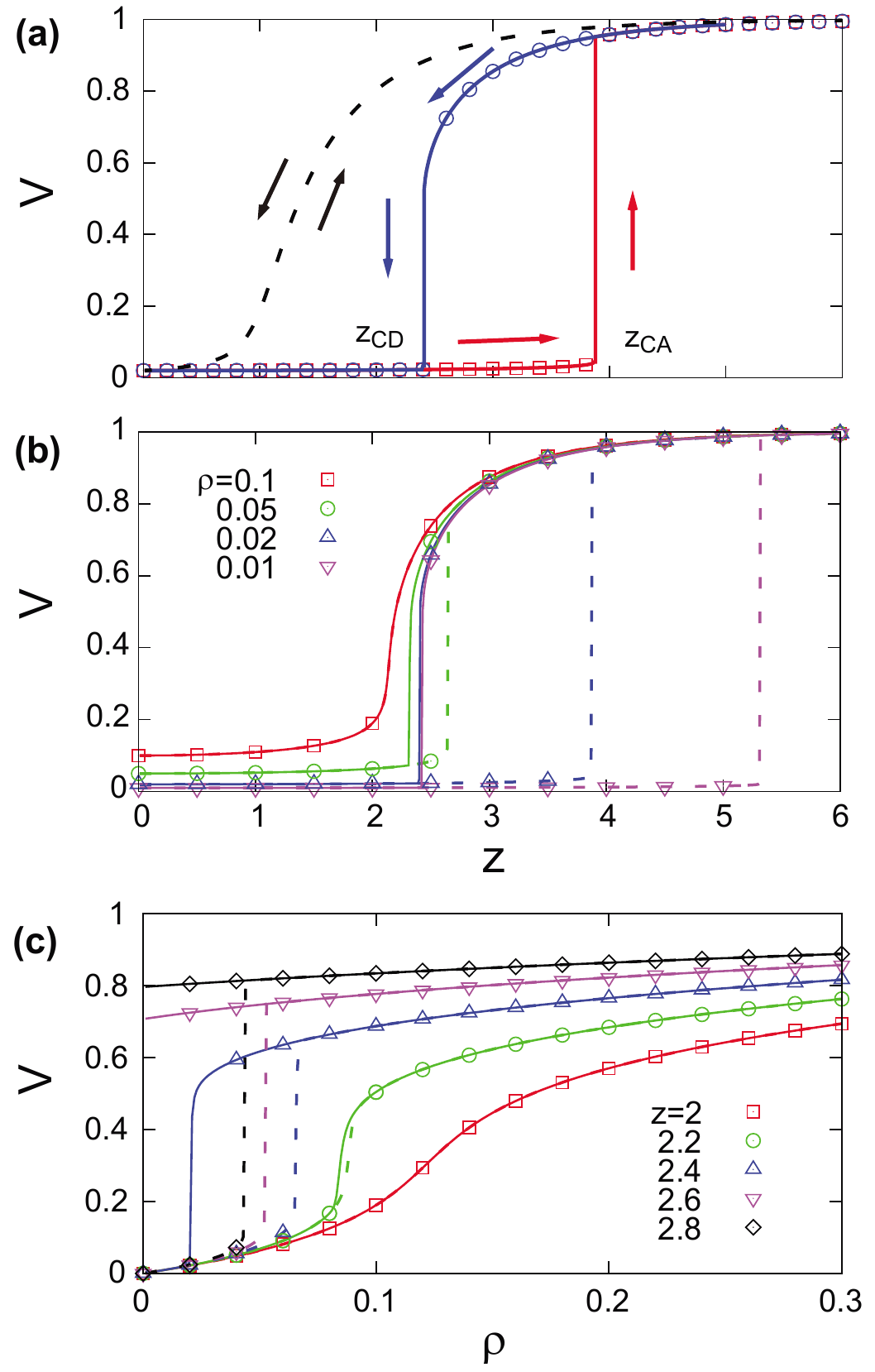}
\caption{
(a) Hysteresis curves with $\rho=0.02$.
Starting from the well-connected high viability state, the systemic collapse ($\circ$) and the subsequent recovery ($\square$) exhibit different curves.
Dashed line indicates viability with $n=1$ for comparison, without hysteresis.
(b) Viability for as a function of $z$ for $\rho=0.1$ $(\square)$, $0.05$ $(\circ)$, $0.02$ $(\triangle)$, and $0.01$ $(\triangledown)$. Note the lack of bistability and hysteresis for $\rho=0.1$.
(c) Viability as a function of $\rho$ for $z=2$ $(\square)$, $2.2$ $(\circ)$, $2.4$ $(\triangle)$, $2.6$ $(\triangledown)$, and $2.8$ $(\diamond)$.
Solid (dashed) lines correspond to the CD (CA) algorithm results.
Both analytic (lines) and numerical results with $N=10^6$ (symbols) are shown together.
}
\end{figure}

It is straightforward to extend to a $n$-resource demand problem on $n$-layer multiplex networks.
For example, in $n$-layer ER networks, $V$ can be obtained by 
\begin{equation}
V=\rho + (1-\rho) (1-e^{-z V})^n.
\end{equation}
The results are qualitatively the same with 2-layer case displaying discontinuity and hysteresis.
The critical point is obtained as 
$(\rho_c,z_c)=(\frac{C_2 \log n}{C_1 +C_2 \log n},\log n +\frac{C_1}{C_2})$,
where $C_1=n^n-(n-1)^n$ and $C_2=n(n-1)^{n-1}$.
$\rho_c$ and $z_c$ increase with $n$ and
$(\rho_c,z_c)\rightarrow (1,\infty)$ in the limit $n \rightarrow \infty$.
Therefore, as $n$ increases, the higher $\rho$ is needed to avoid discontinuity and hysteresis, increasing the system's susceptibility to the abrupt collapse and excessive recovery.
We can also make the problem more realistic that each resource node can supply only one kind of resource, rather than all resources.
For the case of equal $\rho$ for each kind of resource on the 2-layer ER networks, $V$ is simply expressed as the solution of 
\begin{equation}
V=[\rho + (1-\rho) (1-e^{-z V})]^2.
\end{equation}
The results are again qualitatively the same but with different critical point $(\rho_c,z_c)=(0.347\dots,2.702\dots)$,
suggesting the increased potential risk of abrupt collapse on a broader range of system parameters.

Finally, worthwhile to note is the relation between our model and other dynamics including bootstrap percolation, mutual percolation on multiplex networks and cooperative epidemics.
For example, the CA algorithm can be regarded as a variant of bootstrap percolation 
applicable to multiplex networks~\cite{weak,bootstrap}.
When each source node supplies all kind of resources, the CA algorithm is indeed the same as weak bootstrap percolation on multiplex networks
independently proposed in \cite{weak}.
On the other hand, the mutual percolation (equivalently cascading failures in interdependent networks) \cite{buldyrev,son}
corresponds to the CD algorithm of our model in the limit $\rho\rightarrow 0$.
In this aspect, our model can be regarded as a generalization of mutual percolation on multiplex networks.
Importantly, in our model the CA and CD algorithms appear inherently complementary, manifesting themselves as parallel realizations of different stable solutions of a single physical problem to drive the bistability and hysteresis in an organic manner.
In addition, from the perspective of epidemic spreading and social contagion,
the condition of viability can be interpreted as cooperative infection \cite{bizhani,chen} and contagion on multiplex networks \cite{brummitt}.
Therefore, our simple model unifies various percolation and epidemic spreading on 
multiplex networks. 

To summarize, we have introduced and studied a simple model to assess viability of multiplex systems demanding
connections to resources through multiple different channels.
The model exhibits rich phenomenology such as hysteresis and discontinuous jumps in viability with the mean degree and the fraction of resource nodes, and 
the critical point for general $n$-layer of ER networks was obtained.
Our result warns that recovery processes after systemic collapse
can be exceedingly costly due to the hysteresis in viability in multiplex systems.
Directions of further study may include examining of the effect of limited capacity, non-random placement of source nodes and designing strategies for optimal recovery, to mention but a few.

\begin{acknowledgments}
This work was supported by the Basic Science Research Program through an NRF grant funded by MSIP (No. 2011-0014191).
B. M. is also supported by a Korea University Grant.
\end{acknowledgments}


\begin{thebibliography}{99}
\bibitem{rinaldi} S. M. Rinaldi, J. P. Peerenboom, and T. K. Kelly, IEEE Contr. Syst. Mag. {\bf 21}, 11 (2001).
\bibitem{utility} D. E. Kullmann, Mathematics Magazine, {\bf 52}, 299 (1979).
\bibitem{albert} R. Albert, H. Jeong, and A.-L. Barab\'asi, Nature {\bf 406}, 378 (2000).
\bibitem{callaway} D. S. Callaway, M. E. J. Newman, S. H. Strogatz, and D. J. Watts, Phys. Rev. Lett. {\bf 85}, 5468 (2000).
\bibitem{cohen} R. Cohen, K. Erez, D. ben-Avraham, and S. Havlin, Phys. Rev. Lett. {\bf 85}, 4626 (2000).
\bibitem{holme} P. Holme, B. J. Kim, C. N. Yoon, and S. K. Han, Phys. Rev. E {\bf 65}, 056109 (2002).
\bibitem{latora} V. Latora and M. Marchiori, Phys. Rev. E {\bf 71}, 015103 (2005).
\bibitem{rosato} V. Rosato, {\it et al.}, Int. J. Crit. Infrastruct. {\bf 4}, 63 (2008).
\bibitem{buldyrev} S. V. Buldyrev, {\it et al.}, Nature {\bf 464}, 7291 (2010).
\bibitem{dsouza} C. D. Brummitt, R. M. D'Souza, and E. A. Leicht, Proc. Natl. Acad. Sci. USA {\bf 109}, E608 (2012). 
\bibitem{li} W. Li, A. Bashan, S. V. Buldyrev, H. E. Stanley, and S. Havlin, Phys. Rev. Lett. {\bf 108}, 228702 (2012).
\bibitem{bashan} A. Bashan, Y. Berezin, S. V. Buldyrev, and S. Havlin, Nat. Phys. {\bf 9}, 667 (2013).
\bibitem{son} S.-W. Son, {\it et al.}, EPL {\bf 97}, 16006 (2012).
\bibitem{baxter} G. J. Baxter, S. N. Dorogovtsev, A. V. Goltsev, and J. F. F. Mendes, Phys. Rev. Lett. {\bf 109}, 248701 (2012).
\bibitem{weak} G. J. Baxter, S. N. Dorogovtsev, J. F. F. Mendes, and D. Cellai, e-print arXiv:1312.3814v1.
\bibitem{janssen} H. K. Janssen, M. M\"uller, and O. Stenull, Phys. Rev. E {\bf 70} 026114 (2004).
\bibitem{bizhani} G. Bizhani, M. Paczuski, and P. Grassberger, Phys. Rev. E {\bf 86}, 011128 (2012).
\bibitem{chen} L. Chen, F. Ghanbarnejad, W. Cai, and P. Grassberger, EPL {\bf 104}, 50001 (2013).
\bibitem{majdandzic} A. Majdandzic, {\it et al.}, Nat. Phys. {\bf 10}, 34 (2014).
\bibitem{bootstrap} G. J. Baxter, S. N. Dorogovtsev, A. V. Goltsev, and J. F. F. Mendes, Phys. Rev. E {\bf 82}, 011103 (2010).
\bibitem{brummitt} C. D. Brummitt, K.-M. Lee, and K.-I. Goh, Phys. Rev. E {\bf 85}, 045102(R) (2012).
\end{thebibliography}
\end{document}